\documentstyle[12pt,epsf]{article}
\begin{document}


\pagestyle{empty}

\renewcommand{\thefootnote}{\fnsymbol{footnote}}
                                                  

\begin{flushright}
{\small
SLAC--PUB--7117\\
January 1997\\}
\end{flushright}
                
\begin{center}
{\bf\large
Measurement of the $B^+$ and $B^0$ Lifetimes from Semileptonic
Decays\footnote{Work supported in part by the
Department of Energy contract  DE--AC03--76SF00515.}}

\bigskip

The SLD Collaboration
\smallskip

Stanford Linear Accelerator Center, \\
Stanford University, Stanford, CA 94309\\
\medskip

\vspace{2.5cm}

{\bf\large
Abstract }
\end{center}
 
The lifetimes of $B^+$ and $B^0$ mesons are measured using             
a sample of 150,000 hadronic $Z^0$ decays collected by the SLD experiment
at the SLC between 1993 and 1995. The analysis identifies the
semileptonic decays of $B$ mesons and
reconstructs the $B$ meson decay length and charge                           
by vertexing the lepton with a partially reconstructed $D$
meson. This new method results in a sample of
634 (584) charged (neutral) decays with high charge purity.
The ratio of $B^+ : B^0$ decays in the charged (neutral) sample is
3:1 (1:3).
A maximum likelihood fit yields
$\tau_{B^+} =1.61^{+0.13}_{-0.12}($stat$)\pm0.07$(syst) ps,
$\tau_{B^0} =1.56^{+0.14}_{-0.13}\pm0.10$ ps,
and
$\tau_{B^+}/\tau_{B^0} = 1.03^{+0.16}_{-0.14}\pm0.09$. 

\vspace{2cm}

\begin{center}

Submitted to {\sl Physical Review Letters}

\end{center}
\vfill

\renewcommand{\baselinestretch}{2}
\normalsize

\pagebreak
\pagestyle{plain}

%
%
%
  \def\iADEL{$^{(1)}$}
  \def\iBOL{$^{(2)}$}
  \def\iBU{$^{(3)}$}
  \def\iBRUN{$^{(4)}$}
  \def\iUCSB{$^{(5)}$}
  \def\iUCSC{$^{(6)}$}
  \def\iCIN{$^{(7)}$}
  \def\iCSU{$^{(8)}$}
  \def\iCOLO{$^{(9)}$}
  \def\iCOL{$^{(10)}$}
  \def\iFER{$^{(11)}$}
  \def\iFRA{$^{(12)}$}
  \def\iILL{$^{(13)}$}
  \def\iLBL{$^{(14)}$}
  \def\iMIT{$^{(15)}$}
  \def\iMASS{$^{(16)}$}
  \def\iMISS{$^{(17)}$}
  \def\iMOSC{$^{(18)}$}
  \def\iNAG{$^{(19)}$}
  \def\iOREG{$^{(20)}$}
  \def\iPAD{$^{(21)}$}
  \def\iPERU{$^{(22)}$}
  \def\iPISA{$^{(23)}$}
  \def\iRUT{$^{(24)}$}
  \def\iRAL{$^{(25)}$}
  \def\iSOGANG{$^{(26)}$}
  \def\iSOONG{$^{(27)}$}
  \def\iSLAC{$^{(28)}$}
  \def\iTENN{$^{(29)}$}
  \def\iTOH{$^{(30)}$}
  \def\iVAND{$^{(31)}$}
  \def\iWASH{$^{(32)}$}
  \def\iWISC{$^{(33)}$}
  \def\iYALE{$^{(34)}$}
  \def\dead{$^{\dag}$}
  \def\andgen{$^{(a)}$}
  \def\andper{$^{(b)}$}
%
%
\begin{center}
\mbox{K. Abe                 \unskip,\iNAG}
\mbox{K. Abe                 \unskip,\iTOH}
\mbox{T. Akagi               \unskip,\iSLAC}
\mbox{N.J. Allen             \unskip,\iBRUN}
\mbox{W.W. Ash               \unskip,\iSLAC$^\dagger$}
\mbox{D. Aston               \unskip,\iSLAC}
\mbox{K.G. Baird             \unskip,\iMASS}
\mbox{C. Baltay              \unskip,\iYALE}
\mbox{H.R. Band              \unskip,\iWISC}
\mbox{M.B. Barakat           \unskip,\iYALE
% ...
}
\it
  \vskip \baselineskip                   
  \centerline{(The SLD Collaboration)}   
  \vskip \baselineskip                   
%
%
%
  \iADEL
     Adelphi University,
     Garden City, New York 11530 \break
  \iBOL
     INFN Sezione di Bologna,
     I-40126 Bologna, Italy \break
  \iBU
     Boston University,
     Boston, Massachusetts 02215 \break
  \iBRUN
     Brunel University,
     Uxbridge, Middlesex UB8 3PH, United Kingdom \break
  \iUCSB
     University of California at Santa Barbara,
     Santa Barbara, California 93106 \break
  \iUCSC
     University of California at Santa Cruz,
     Santa Cruz, California 95064 \break
  \iCIN
     University of Cincinnati,
     Cincinnati, Ohio 45221 \break
  \iCSU
     Colorado State University,
     Fort Collins, Colorado 80523 \break
  \iCOLO
     University of Colorado,
     Boulder, Colorado 80309 \break
  \iCOL
     Columbia University,
     New York, New York 10027 \break
  \iFER
     INFN Sezione di Ferrara and Universit\`a di Ferrara,
     I-44100 Ferrara, Italy \break
  \iFRA
     INFN  Lab. Nazionali di Frascati,
     I-00044 Frascati, Italy \break
  \iILL
     University of Illinois,
     Urbana, Illinois 61801 \break
  \iLBL
     Lawrence Berkeley Laboratory, University of California,
     Berkeley, California 94720 \break
  \iMIT
     Massachusetts Institute of Technology,
     Cambridge, Massachusetts 02139 \break
  \iMASS
     University of Massachusetts,
     Amherst, Massachusetts 01003 \break
  \iMISS
     University of Mississippi,
     University, Mississippi  38677 \break
  \iMOSC
    Moscow State University,
    Institute of Nuclear Physics
    119899 Moscow, Russia    \break
  \iNAG
     Nagoya University,
     Chikusa-ku, Nagoya 464 Japan  \break
  \iOREG
     University of Oregon,
     Eugene, Oregon 97403 \break
  \iPAD
     INFN Sezione di Padova and Universit\`a di Padova,
     I-35100 Padova, Italy \break
  \iPERU
     INFN Sezione di Perugia and Universit\`a di Perugia,
     I-06100 Perugia, Italy \break
  \iPISA
     INFN Sezione di Pisa and Universit\`a di Pisa,
     I-56100 Pisa, Italy \break
  \iRUT
     Rutgers University,
     Piscataway, New Jersey 08855 \break
  \iRAL
     Rutherford Appleton Laboratory,
     Chilton, Didcot, Oxon OX11 0QX United Kingdom \break
  \iSOGANG
     Sogang University,
     Seoul, Korea \break
  \iSOONG
     Soongsil University,
     Seoul, Korea  156-743 \break
  \iSLAC
     Stanford Linear Accelerator Center, Stanford University,
     Stanford, California 94309 \break
  \iTENN
     University of Tennessee,
     Knoxville, Tennessee 37996 \break
  \iTOH
     Tohoku University,
     Sendai 980 Japan \break
  \iVAND
     Vanderbilt University,
     Nashville, Tennessee 37235 \break
  \iWASH
     University of Washington,
     Seattle, Washington 98195 \break
  \iWISC
     University of Wisconsin,
     Madison, Wisconsin 53706 \break
  \iYALE
     Yale University,
     New Haven, Connecticut 06511 \break
  \dead
     Deceased \break
  \andgen
     Also at the Universit\`a di Genova \break
  \andper
     Also at the Universit\`a di Perugia \break
\rm
%

\end{center}


\pagebreak
 

According to the spectator model of heavy hadron weak decay, the heavy
quark decays independently of the other quarks in the hadron.
Therefore, this model predicts that the lifetimes of all hadrons
containing a given heavy quark Q are equal.
However, the hierarchy observed in the charm system,
$\tau_{D^+} > \tau_{D_s^+} \sim \tau_{D^0} > \tau_{\Lambda_c^+}$, indicates
the need for corrections to this model.
Such lifetime differences are predicted to scale with $1/m_Q^2$~\cite{bigi}
and therefore are expected to be less than 10\% in the $b$-quark system.
Measurements of the $B^+$ and $B^0$ lifetimes
provide tests of this prediction.
Finally, precise measurements of exclusive $B$ meson lifetimes
are necessary to extract the CKM matrix element $V_{cb}$.

The measurements presented here use a sample of
150,000 hadronic $Z^0$ decays collected between 1993 and 1995
by the SLD experiment at the SLC.
The analysis uses a new technique
to identify the $B$ hadron charge by
reconstructing the charged track topology
of both $B$ and cascade $D$ vertices in semileptonic $B$ decays.
Most previous measurements of the $B^+$ and $B^0$ lifetimes~\cite{Semil}
are based on samples of semileptonic decays in which the lepton is
identified and the $D^{(\ast)}$ meson is fully reconstructed.
The lifetime measurements then rely on assumptions concerning the
$B^+$ and $B^0$ content of the $\overline{D^0} X l^+\nu$ and
$D^{(\ast)-} X l^+\nu$ samples.
In contrast, the more inclusive technique used here
only relies on the simple difference of total charge between
$B^+$ and $B^0$ decays.
However, this technique requires very good vertexing
to assign tracks correctly to secondary vertices.

The analysis uses the calorimetry and tracking systems
(for details see Ref.~\cite{rbrb}).
The Liquid Argon Calorimeter (LAC) is used to reconstruct jets from
energy clusters and perform electron identification
with maximal efficiency for $|\cos\theta|<0.72$.
The Warm Iron Calorimeter (WIC) provides efficient muon identification
for $|\cos\theta|<0.60$.
Tracking is performed with the Central Drift Chamber (CDC) and the
CCD pixel Vertex Detector (VXD) with 94\% total reconstruction
efficiency for $\left|\cos\theta\right|<0.74$ (including VXD hit linking).
The impact parameter resolution for high-momentum tracks is measured using
$Z^0\rightarrow\mu^+\mu^-$ decays to be 11~$\mu$m in the
plane perpendicular to the beam axis ($r\phi$ plane) and 38~$\mu$m
in the plane containing the beam axis ($rz$ plane).

The decay length is measured relative to the
position of the micron-size SLC Interaction Point (IP) which is
reconstructed in the $r\phi$ plane with a
precision of $\sigma_{r\phi} = (7\pm2)\:\mu$m using tracks in sets of
$\sim30$ sequential hadronic $Z^0$ decays.
The $z$ position of the IP is determined on an
event-by-event basis with $\sigma_z \simeq 52\:\mu$m
for $b\:\bar{b}$ events~\cite{rbrb}
using the median $z$ position of tracks at their
point-of-closest-approach to the IP in the $r\phi$ plane.
 
The measurements rely on a Monte Carlo simulation based on the
JETSET~7.4 event generator~\cite{Jetset}
and the GEANT~3.21 detector simulation package~\cite{Geant}
to determine the charge separation purity and to extract the
lifetimes from the decay length distributions.
The $b$-quark fragmentation follows the Peterson {\it et al.}
parameterization~\cite{Peterson}.
$B$ mesons (baryons) are generated with mean lifetime
$\tau = 1.55$ ps ($\tau = 1.10$ ps).
$B$ meson decays are modelled according to the
CLEO $B$ decay model \cite{BdkModel} tuned to reproduce
the spectra and multiplicities
of leptons, charmed hadrons, pions, kaons, and protons,
measured at the $\Upsilon$(4S)~\cite{argcl}.
Semileptonic decays follow the ISGW model~\cite{ISGW}
including 23\% $D^{\ast\ast}$ production.
$B$ baryon and charmed hadron decays are modelled using JETSET with,
in the latter case, branching fractions tuned to
existing measurements~\cite{PDG94}.

The initial step in the event selection is to select
electron and muon candidates using the measured
track parameters as well as measurements from the LAC and WIC
respectively (see Ref.~\cite{Ablept} for further details).
To enhance the fraction of $Z^0\rightarrow b\:\bar{b}$ events with
little loss in efficiency,
lepton candidates are required to have
total momentum $p > 2$ GeV/c and momentum transverse
to the nearest jet $> 0.4$ GeV/c (jets are found using the JADE
algorithm~\cite{JADE} with $y_{cut} = 0.005$).
These cuts yield a sample of $\sim34,000$ event hemispheres,
with an efficiency
of $\sim 75\%$ for semileptonic $B$ decays within $|\cos\theta|<0.6$
determined from our Monte Carlo simulation.

The secondary vertex reconstruction proceeds separately
for each event hemisphere containing a lepton, and uses a multi-pass
algorithm that operates on
those tracks that have at least one VXD hit and are not
from identified $\gamma$ conversions, 
or $K^0_s$ or $\Lambda$ decays.
Tracks are initially classified as primary unless their 3-D
impact parameter significance with respect to the IP is $> 3.5\:\sigma$ and
$p > 0.8$ GeV/c, in which case they are classified as secondary.

In the first pass, the hemisphere containing the lepton candidate
is required to include no more than four secondary tracks
(excluding the lepton) and a candidate $D$ vertex is constructed
using all such tracks (vertex cuts are defined below).
The $D$ trajectory, found from the $D$ vertex and the total momentum
vector of tracks included in the vertex, must intersect the lepton
to form a valid one-prong $B$ vertex solution.
If this is successful, an attempt is made to form a two-prong
$B$ vertex by attaching one primary track to the lepton near the point
of intersection.
This first pass identifies $91\%$ of the final candidates.  
These candidates are allowed to be modified by
searching for one or two primary tracks that can be added to the 
existing $D$ vertex.
This search is successful for $40\%$ of the candidates.
In case of multiple solutions, we select the one with the
largest number of tracks and if more than one still remains,
we select that with the smallest impact
parameter between the $D$ trajectory and the lepton or two-prong $B$ vertex.
A second pass is performed if no first pass candidate is identified.
Here, a search is made for solutions in which one
secondary track makes a valid two-prong $B$ vertex with the lepton,
the remaining secondary tracks form a $D$ vertex,
and the $D$ trajectory intersects the $B$ vertex.
Multiple solutions are handled as described above.

The requirements to form a $D$ vertex are:
the number of tracks is $\leq 4$;
the absolute value of the charge  $\leq 1$;
the mass (charged tracks assumed to be $\pi$'s) $< 1.98$ GeV/c$^2$;
the vertex displacement from the IP $> 4\:\sigma$ and $< 2.5$ cm; and
the vertex $\chi^2$ (2,3,4 prongs) $< (4, 12, 20)$.
The requirements to form a $B$ vertex are:
the absolute value of the total charge ($B$+$D$ tracks) $\leq 1$;
the mass $>$ 1.4 GeV/c$^2$;
the observed decay length $> 0.08$ cm
and $< 2.4$ cm; and
the momentum of the non-lepton track (if any) $> 0.4$ GeV/c.
The requirements for the $D$ vertex to be linked to the $B$ vertex are:
the signed distance between $D$ and $B$ vertices $> 200~\mu$m;
for one-prong $B$ vertices, the distance of closest approach of the
$D$ trajectory to the lepton $< (130, 100, 70)~\mu$m for
(2, 3, 4) prong $D$ vertices;
for two-prong $B$ vertices, the three-dimensional impact parameter of
the $D$ trajectory with respect to the $B$ vertex $< 200~\mu$m.

The algorithm yields
783 charged and 584 neutral semileptonic $B$ decay candidates.
The topological breakdown is given in Table~\ref{leptopol}.
The efficiency for reconstructing a semileptonic $B$ decay
is estimated from the simulation to be 24\% for decays with
an identified lepton within $|\cos\theta|<0.6$.
 
\begin{table}[htb]
\begin{center}                                                                
\caption{\label{leptopol} Summary of reconstructed topologies, including
         the fraction of each topology in the combined charged and neutral
         sample for data and Monte Carlo simulation.}
\begin{tabular}{lcccrr}
\\
\hline
           & $B$ Vertex & $D$ Vertex & \multicolumn{2}{c}{Data} & MC \\
           &            &            & \multicolumn{1}{c}{\# decays}
           &  \multicolumn{1}{c}{Fraction} & Fraction \\
\hline                                                                        
           & 1 prong    & 2 prong & 519 & $(38.0\pm1.3)\%$ & 37.6\% \\
 $Q=\pm 1$ & 1 prong    & 4 prong & 115 &  $(8.4\pm0.8)\%$ &  8.5\% \\
           & 2 prong    & 3 prong & 149 & $(10.9\pm0.8)\%$ &  9.6\% \\
\hline                                          
           & 1 prong    & 3 prong & 341 & $(24.9\pm1.2)\%$ & 26.8\% \\
 $Q=0$     & 2 prong    & 2 prong & 175 & $(12.8\pm0.9)\%$ & 13.6\% \\ 
           & 2 prong    & 4 prong & ~68 &  $(5.0\pm0.6)\%$ &  3.9\% \\ 
\hline
\end{tabular}                                                                 
\end{center}
\end{table}

Monte Carlo studies indicate that the $B^+$ topology consisting of
two-prong $B$ and three-prong $D$ vertices has poor $B^+$ purity
due to the small $B^+ \rightarrow D^-\pi^+ l^+\nu$ branching ratio and
the large background from  $B^0 \rightarrow D^{(\ast)-} l^+\nu$ decays.
This is corroborated by the large fraction of decays
with $B$ vertex charge = 0 observed in the data for this topology.
Therefore, this topology is rejected thereby reducing the charged sample
to 634 candidates. These studies show that
the remaining charged (neutral) sample is 97.4\% (98.9\%) pure in
$B$ hadrons with flavor contents of 66.6\% $B^+$,
22.9\% $B^0$, 5.5\% $B_s^0$, and 2.4\% $B$ baryons for the charged
sample, and
19.5\% $B^+$, 60.2\% $B^0$, 14.8\% $B_s^0$, and 4.4\% $B$ baryons
for the neutral sample.
The sensitivity of the analysis to the individual $B^+$ and
$B^0$ lifetimes can be assessed from the 3:1 ratio of
$B^+\:(B^0)$ over $B^0\:(B^+)$ decays
in the charged (neutral) sample.
The fraction of misidentified leptons is
7.0\% (9.7\%) for charged (neutral) candidates,
as determined from the simulation.
However, these are predominantly $B$ decays with good charge purity.

\begin{figure}[p]
\centering
\vspace{-0.8truecm}
\centering
\epsfysize14.0cm
\leavevmode
\epsfbox{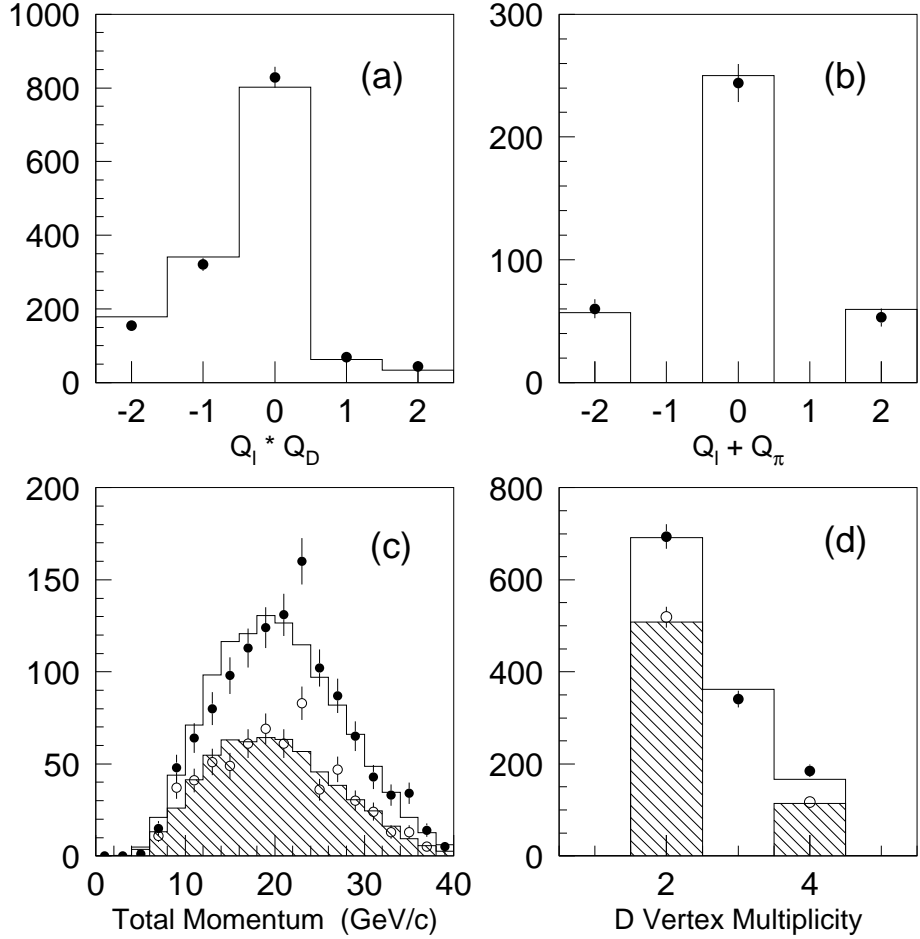}
\caption{\label{qcomplep} Distributions of (a) the product of lepton and
         $D$-vertex charges, (b) sum of lepton and slow transition pion
         charges for data (points) and Monte Carlo simulation
         (histograms) with no charge requirement at the $B$ and $D$ vertices.
         Distributions of (c) total momentum of the $B$+$D$ tracks
         and (d) $D$ vertex multiplicity
         for data (solid circles are for total sample,
         open circles are for charged sample only)
         and Monte Carlo simulation (histograms are for total sample,
         shaded portions are for charged sample only).}
\end{figure}

As a check of the algorithm, the requirements on
the charges of the $B$ and $D$ vertices are removed
for Figs.~\ref{qcomplep}(a) and \ref{qcomplep}(b).
Figure~\ref{qcomplep}(a) shows that, as expected, the charges of
the lepton and $D$ vertex are opposite for most reconstructed decays
(provided the $D$ vertex is charged).
Furthermore, the charge distribution resulting from
the lepton$+$slow transition pion vertex (from $D^{\ast(\ast)}$)
shown in Fig.~\ref{qcomplep}(b)
indicates that the track combined with the      
lepton to form a two-prong $B$ vertex most often has charge opposite that
of the lepton, as expected for $B\rightarrow D^{\ast}l\nu$
and most $B\rightarrow D^{\ast\ast}l\nu$ decays.
Figure~\ref{qcomplep} also shows the total vertex momentum
distribution obtained using the tracks from both $B$ and $D$ vertices,
and the $D$ vertex multiplicity distribution
(with the nominal charge requirements on the vertices).
Overall, there is agreement between the data and the Monte Carlo
simulation.
 
\begin{figure}[htb]
\centering
\epsfysize9.7cm
\leavevmode
\epsfbox{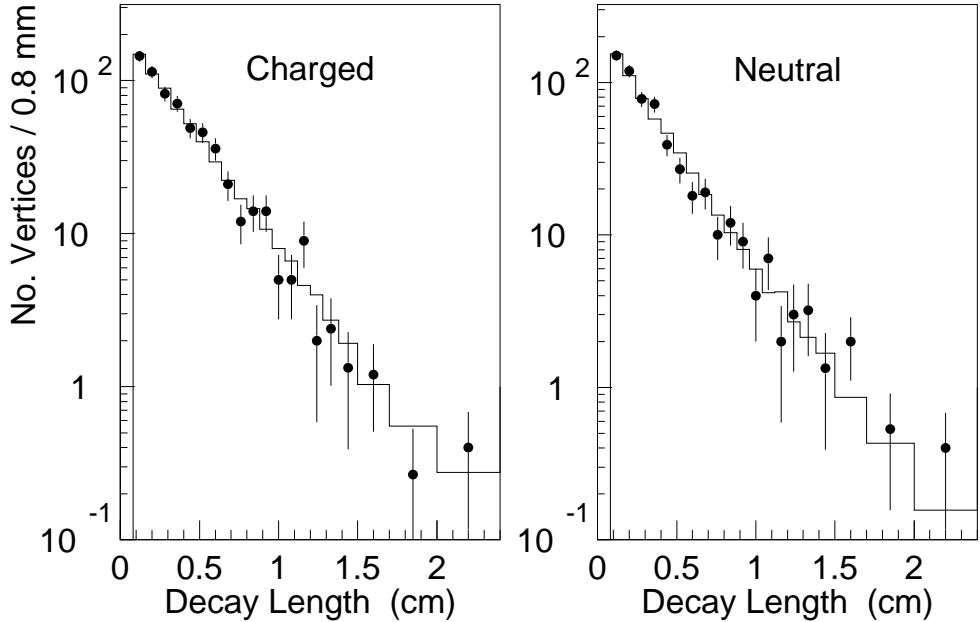}
\caption{Decay length distributions for charged and neutral decays
         for data (points) and the Monte Carlo simulation corresponding
         to the best fit (histograms).}
\label{semidklen}                                                             
\end{figure}

The $B^+$ and $B^0$ lifetimes are extracted from the decay length
distributions of the $B$ vertices in the charged and neutral
samples (see Fig.~\ref{semidklen})
using a binned maximum likelihood technique.
These distributions are fitted
simultaneously to determine the $B^+$ and $B^0$ lifetimes.
For each set of parameter values, Monte Carlo decay length distributions
are obtained by reweighting entries from generated $B^+$ and $B^0$
decays in the original Monte Carlo decay length             
distributions with
$W(t,\tau)= \left( \frac{1}{\tau} ~ e^{-t/\tau} \right)/
 \left( \frac{1}{\tau_{gen}} ~ e^{-t/\tau_{gen}} \right)$,
where $\tau$ is the desired $B^+$ or $B^0$ lifetime,                          
$\tau_{gen}= 1.55$ ps, and $t$ is the proper time of each decay.           
The fit then compares the decay length distributions
from the data with the reweighted Monte Carlo distributions.
The fit yields
\begin{eqnarray}
\tau_{B^+} & = & 1.61^{+0.13}_{-0.12} \mbox{~ps,} \nonumber \\
\tau_{B^0} & = & 1.56^{+0.14}_{-0.13} \mbox{~ps,} \nonumber \\
\tau_{B^+}/\tau_{B^0} & = & 1.03^{+0.16}_{-0.14} . \nonumber
\end{eqnarray}
with a $\chi^2 = 78$ for 76 degrees of freedom.
  
Systematic uncertainties due to detector and
physics modeling, as well as those related to the fitting procedure,          
are described below and summarized in Table~\ref{systerrs}.
A discrepancy between data and simulation in the fraction
of tracks passing a set of quality cuts~\cite{rbrb}
is corrected for by removing 4\% of the tracks from
the simulation.
In addition, a $0.9^{m-2}$ correction
to the vertex reconstruction efficiency is applied,
where $m$ is the $D$ vertex track multiplicity.
The uncertainty due to the track finding
efficiency is conservatively
estimated as the full difference between fits with and without
these corrections.
\begin{table}[hb]
\begin{center}                                                                
\caption{\label{systerrs} Summary of systematic
         uncertainties in the $B^+$ and $B^0$ lifetimes and their ratio.}
\begin{tabular}{lcccc}
\\
\hline
Systematic Error &  & $\Delta\tau_{B^+}$
                    & $\Delta\tau_{B^0}$
                    & $\Delta\frac{\tau_{B^+}}{\tau_{B^0}}$ \\
 & & (ps) & (ps) & \\
\hline
 & \multicolumn{4}{c}{Detector Modeling} \\ \hline
Tracking efficiency      &  & 0.017 & 0.029 & 0.023 \\
Tracking resolution      &  & 0.020 & 0.030 & 0.033 \\
Lepton misidentif.       &  & 0.006 & 0.007 & $<$.003 \\ \hline
 & \multicolumn{4}{c}{Physics Modeling} \\ \hline
$b$ fragmentation        & $0.700 \pm 0.011$
                         & 0.035 & 0.039 & 0.016 \\
$\frac{BR(B\rightarrow D^{**}l\nu)}{BR(B\rightarrow X l\nu)}$
                         & $0.230 \pm 0.115$
                         & 0.011 & 0.018 & 0.016 \\
BR($B\rightarrow D \overline{D} X$) & $0.15 \pm 0.05$
                         & 0.009 & 0.008 & 0.011 \\
$B_s^0$ fraction         & $0.115 \pm 0.040$
                         & 0.007 & 0.007 & 0.009 \\
$B$ baryon fraction      & $0.072 \pm 0.040$
                         & 0.008 & 0.016 & 0.006 \\
$B_s^0$ lifetime         & $1.55 \pm 0.10$ ps
                         & 0.003 & 0.028 & 0.020 \\
$B$ baryon lifetime      & $1.10 \pm 0.08$ ps
                         & $<$.003 & 0.007 & 0.005 \\
$D$ decay multipl.       & Ref.\cite{MK3DDECAY}
                         & 0.014 & 0.009 & $<$.003 \\
$D$ mom. mismatch     &  & $<$.003 & 0.034 & 0.022 \\
 \hline
 & \multicolumn{4}{c}{Monte Carlo and Fitting} \\ \hline
Fitting systematics   &  & 0.037 & 0.052 & 0.061 \\
MC statistics         &  & 0.023 & 0.024 & 0.027 \\
\hline
\hline
TOTAL                 &  & 0.066 & 0.097 & 0.088 \\
\hline
\end{tabular}                                                                 
\end{center}                                                       
\end{table}
The uncertainty due to tracking resolution is similarly taken
to be the difference between fits before and after smearing and shifting
the track impact parameters in the $rz$ plane
to account for residual VXD misalignments~\cite{rbrb}.
The smearing
by $\sigma = 20\,\mu$m/~sin$\,\theta$ and shifts
up to 20 $\mu$m are required to match the core of the impact
parameter distribution observed in the data.
No correction is required to the impact parameters in the $r\phi$ plane.
The $B^0$ lifetime is more sensitive to the above uncertainties than
the $B^+$ lifetime because they affect the relative abundance of
the various topologies (listed in Table~\ref{leptopol}) and
the amount of wrong-charge vertices at short decay length is higher for
two-prong than for one-prong $B$ vertex topologies.
The rate of lepton misidentification is varied by
$\pm$25\% in the simulation.
It was checked that the lifetimes obtained in four different
azimuthal regions are statistically consistent.

The $b$-quark fragmentation uncertainty includes contributions
from shifting the mean value of the $B$ hadron energy~\cite{bfrag}
and using a different fragmentation function shape~\cite{Bowler}.
As expected, this uncertainty affects the individual lifetimes but
leaves the lifetime ratio relatively unaffected.
Uncertainties in the $B_s^0$ and $B$ baryon production and lifetimes
contribute more significantly to the $B^0$ lifetime,
and thus affect the lifetime ratio,
due to the larger fraction of $B_s^0$ and $B$ baryons in the
neutral sample.
Sensitivity to the branching ratio for decays involving
$b\rightarrow c\rightarrow l$ transitions or
for $B\rightarrow \tau \nu_\tau X$ decays is negligible.
Similarly, uncertainties due to the charmed hadron lifetimes are negligible.

A slight discrepancy between data and simulation is observed
in the vertex total momentum distribution for the neutral sample
(see Fig.~\ref{qcomplep}(c)).
This mismatch is investigated by 
reweighting the Monte Carlo $D$ vertex momentum distribution
to match the data in both charged and neutral samples.
Although the discrepancy
may be attributed in part to the $B$ decay modeling,
we conservatively assign an uncertainty
to be the difference between fits with and without reweighting.

The fitting uncertainties are estimated
by varying the bin size and the minimum and maximum decay length cuts.
Although the lifetimes obtained for each of these
variations are statistically consistent, we
conservatively assign an uncertainty equal to the root mean
square value of all these results. This uncertainty dominates all others
but is largely driven by the available statistics.

The final results are
\begin{eqnarray}
\tau_{B^+} & = & 1.61^{+0.13}_{-0.12}(\mbox{stat})\pm0.07(\mbox{syst})\mbox{~ps}, \nonumber \\
\tau_{B^0} & = & 1.56^{+0.14}_{-0.13}(\mbox{stat})\pm0.10(\mbox{syst})\mbox{~ps}, \nonumber \\
\tau_{B^+}/\tau_{B^0} & = & 1.03^{+0.16}_{-0.14}(\mbox{stat})\pm0.09(\mbox{syst}). \nonumber
\end{eqnarray}
These results complement those obtained with an inclusive topological
technique~\cite{SLDtopol},
and are in agreement with previous measurements~\cite{Semil,excinc} and
with the expectation that the $B^+$ and $B^0$ lifetimes are nearly equal.

\noindent

We thank the personnel of the SLAC accelerator department and
the technical staffs of our collaborating institutions for their outstanding
efforts.

\end{document}